\documentclass[acmsmall]{acmart}
\AtBeginDocument{%
  }

\setcopyright{acmlicensed}
\copyrightyear{2018}
\acmYear{2018}
\acmDOI{XXXXXXX.XXXXXXX}

\acmJournal{JACM}
\acmVolume{37}
\acmNumber{4}
\acmArticle{111}
\acmMonth{8}

\begin{document}

% \title{Can We Really Explain Deep Foundation Models?}
% \title{The Fundamental Challenges of Explaining Deep Foundation Models}
\title{Intrinsic Barriers to Explaining Deep Foundation Models}

\author{Zhen Tan}
\email{ztan36@asu.edu}
\orcid{0009-0006-9548-2330}
\affiliation{%
  \institution{Arizona State University}
  \city{Tempe}
  \state{Arizona}
  \country{USA}
}

\author{Huan Liu}
\email{huanliu@asu.edu}
\orcid{0000-0002-3264-7904}
\affiliation{%
  \institution{Arizona State University}
  \city{Tempe}
  \state{Arizona}
  \country{USA}
}

\renewcommand{\shortauthors}{Tan et al.}

% \begin{abstract}
%   Deep Foundation Models (DFMs), encompassing systems like Large Language Models, are rapidly reshaping technology and society, demonstrating remarkable capabilities learned from vast datasets. Their integration into diverse and critical applications fuels a strong societal and scientific desire to understand their internal workings and decision-making processes. Yet, these models, characterized by unprecedented scale and complexity, often operate in ways that resist straightforward interpretation. Questions arise about the true depth of understanding we can achieve, given their emergent properties and the apparent limitations of current analytical tools. Can we genuinely unravel the intricate logic hidden within billions of parameters? This paper embarks on a critical reflection of the obstacles confronting the explainability of DFMs. We delve into the fundamental challenges rooted in the intrinsic nature of these complex systems and critically assess the capabilities and limitations of contemporary explainability methods, probing the boundaries of our current ability to render these powerful models transparent.

% \end{abstract}

\begin{abstract}
   Deep Foundation Models (DFMs) offer unprecedented capabilities but their increasing complexity presents profound challenges to understanding their internal workings – a critical need for ensuring trust, safety, and accountability. As we grapple with explaining these systems, a fundamental question emerges: Are the difficulties we face merely temporary hurdles, awaiting more sophisticated analytical techniques, or do they stem from \emph{intrinsic barriers} deeply rooted in the nature of these large-scale models themselves? This paper delves into this critical question by examining the fundamental characteristics of DFMs and scrutinizing the limitations encountered by current explainability methods when confronted with this inherent challenge. We probe the feasibility of achieving satisfactory explanations and consider the implications for how we must approach the verification and governance of these powerful technologies.
\end{abstract}





\maketitle

\section*{Introduction}

Deep Foundation Models (DFMs) – such as large language models and multimodal architectures – are a class of neural networks trained on vast amounts of data, designed to serve as general-purpose engines for downstream tasks across diverse domains~\cite{zhou2024comprehensive}. With the emergence of systems like GPT, Gemini, and CLIP, artificial intelligence is undergoing a profound transformation. These models exhibit remarkable, often unexpected capabilities – from language generation and translation to scientific reasoning and creative synthesis – learned with immense and diverse datasets.
% Systems like GPT-4, PaLM 2, Claude, and their contemporaries demonstrate remarkable, often surprising, capabilities across a vast range of domains, learned from exposure to immense datasets \cite{llm_capabilities_placeholder}. 
As these models are increasingly deployed in science, industry, and everyday applications – influencing decisions, generating content, and mediating interactions – a critical question emerges with growing urgency: Can we understand how they work?

% The desire for explaining their behaviors is not merely academic curiosity; it stems from fundamental needs for reliability, safety, fairness, and accountability. How can we trust a DFM's output in high-stake situations without understanding its internal workings? How can we debug failures, audit for biases, or ensure alignment with human values if the underlying processes remain opaque? The societal and scientific demand for explainable AI (XAI) for DFM is understandably strong.

The desire for explaining DFM behaviors is not merely academic curiosity; it stems from fundamental needs for trust, safety, and accountability. Explanations, for instance, are expected to help build trust in DFM outputs, especially in high-stake situations, by elucidating decision-making steps. They are seen as essential tools to aid in debugging failures, auditing for biases, and ensuring alignment with human values, by uncovering potential erroneous steps. The societal and scientific demand for explainable AI (XAI) for DFMs is understandably strong.

Yet, DFMs present an unprecedented challenge to this demand. Unlike many previous engineered systems, their complexity is not merely a matter of intricate design, but often is a ingrained characteristic arising from their very construction and learning paradigm. They operate at scale and exhibit behaviors that seem to defy mechanistic interpretation using the state-of-the-art tools. This raises a crucial question: Are the difficulties we face in explaining DFMs merely temporary hurdles, awaiting better techniques, or do they stem from \emph{intrinsic barriers} from within the models themselves?

We argue for the latter perspective. We contend that there exist intrinsic obstacles to explaining DFMs. Our focus here is to dissect these inherent barriers by expounding three aspects:
\textcolor{black}{
(1) The sheer \emph{scale and dimensionality} of these models.
(2) Their inherent \emph{non-linearity and sensitivity}.
(3) The unidentifiable \emph{data influence}.}
For each barrier, we will examine why it poses a fundamental challenge and how it manifests in the limitations of current attempts at explanation. 
By focusing on these intrinsic properties, we aim to foster a deeper understanding of why explaining DFMs is so profoundly difficult, and provide implications of the findings.

\section*{The Barrier of Scale and Dimensionality}

% Consider the sheer \emph{scale and dimensionality} of contemporary Deep Foundation Models (DFMs). We routinely encounter systems with hundreds of billions, even trillions, of parameters interacting within vast, high-dimensional activation spaces. What does this immense scale imply for explainability? It raises questions about the limits of both human comprehension and computational explanantion.
% How can we hope to map or truly understand a system whose state space dwarfs that of traditional engineered systems by orders of magnitude?

Consider the sheer \emph{scale and dimensionality} of contemporary Deep Foundation Models (DFMs). We routinely encounter systems with hundreds of billions, even trillions, of parameters interacting within vast, high-dimensional activation spaces. What does this immense scale imply for explainability? It raises questions about the capacity limit of both human comprehension and computational explanations: how can the effectively \textit{infinite} combinatorial state space of such a system be meaningfully mapped onto \textit{finite} human concepts or explanatory models?

This staggering scale directly impacts the feasibility and reliability of current explainability methods. For instance, \emph{feature attribution} techniques (like LIME~\cite{ribeiro2016should} or SHAP~\cite{lundberg2017unified}) often aim to assign importance scores to input features. But how meaningful are such attributions when dealing with potentially billions of input tokens, features, or dimensions interacting in complex ways? Can local, instance-specific analyses provide a sufficiently rich picture of global behavior across trillions of parameters? Furthermore, the computational cost of applying many sophisticated XAI techniques grows significantly with model size, potentially rendering thorough analysis impractical. Similarly, efforts to understand model internals by probing individual \emph{neuron activations} face a combinatorial explosion; assigning clear semantic roles becomes exceptionally difficult when dealing with billions of units, many potentially polysemantic.

Critically, many explanation methods rely on surrogate models or analytical tools that are orders of magnitude smaller and simpler than the DFMs they aim to interpret~\cite{sigma2019interpretability}. Whether through \emph{linear probes}, interpretable approximations like rule lists or decision trees, or techniques like \emph{model distillation}, these approaches attempt to compress or approximate vast, high-dimensional processes into lower-dimensional narratives. This fundamental scale mismatch introduces profound epistemic limitations: can a system with orders of magnitude fewer parameters—or indeed, a human analyst constrained by cognitive limits—truly and faithfully explain the complex, potentially emergent behavior of a system operating at the trillion-parameter scale?

One might counter that some emerging techniques attempt to circumvent this specific scale mismatch by employing \emph{larger DFMs themselves as ``explainers''} to interpret the behavior or outputs of smaller, target DFMs~\cite{Casper2023ExplainingNeurons,tan2024interpreting}. While this approach avoids the problem of explaining a complex system with a vastly simpler one, does it truly resolve the underlying barrier posed by scale? It seemingly only displaces the challenge: we are now faced with the recursive problem of explaining the even larger ``explainer'' DFM. If the explainer itself operates at a scale that defies deep comprehension according to the arguments above, have we made fundamental progress, or merely shifted the locus of inscrutability one level higher? This suggests the core issue of scale persists, demanding justification for the explainer model itself.

The sheer scale of DFMs, therefore, creates a fundamental barrier. It pushes many current explanation approaches beyond their practical limits, raises deep questions about the epistemic validity of reducing complexity via simpler models, and even challenges approaches using larger models by introducing a recursive explanation problem. Scale fundamentally prohibits our ability to achieve human-understandable accounts of these vast systems.

\section*{The Barrier of Non-Linearity and Sensitivity}

Beyond their scale, DFMs exhibit profound \emph{non-linearity and sensitivity}, rooted in their architectural and training dynamics. Deep stacks of non-linear transformations give rise to complex input-output mappings, where small perturbations—whether in input phrasing, internal states, or even the training process—can cascade into significantly different outcomes. This fragility is not incidental; it is fundamental to how DFMs learn and generalize. For instance, large models trained on the same dataset but with different random initializations or data orders can arrive at divergent internal representations~\cite{jordan2024on}.

An example of this phenomenon is the so-called \emph{Reversal Curse}~\cite{berglund2024the}. Models that confidently assert “A + B = C” can fail to infer “C = A + B”, despite the logical equivalence. This reflects an overreliance on surface-level patterns and directional associations in training data, rather than an internalization of symmetric logical structures. Such failures highlight how non-linear learning and path-dependencies, rather than logical consistency, shape the model’s representations. In this context, even slight variations in phrasing can produce qualitatively different outputs, raising a key question: How can we hope to explain behavior in a system whose decision boundaries are both opaque and highly unstable?

This sensitivity poses a major challenge for current XAI techniques, most of which operate under \emph{deterministic} and \emph{static} assumptions. Given a model and an input, methods such as SHAP, LIME, or gradient-based saliency typically yield one fixed explanation. Yet DFMs are dynamic in nature: their outputs may hinge on subtle nuances, context shifts, or historical quirks in training data. How can static, deterministic methods capture the behavior of systems that are not only non-linear but also deeply context-sensitive? If a model behaves differently due to imperceptible variations, how can a single, pointwise attribution claim to reflect “why” a model responded as it did?
This mismatch raises concerns about \emph{faithfulness}: if an explanation method fails to account for the model’s nonlinear dependencies and sensitivity, can it truly reflect the reasons behind a model’s decisions—especially for subtle or counterintuitive behaviors? Worse still, such tools might offer misleadingly coherent narratives for behaviors that are, in reality, unstable or accidental.

Compounding this issue, non-linearity at scale often gives rise to \emph{emergent behavior}—capabilities such as in-context learning or multi-step reasoning that arise suddenly during training of the DFMs, without explicit architectural changes~\cite{wei2022emergent}. These capabilities are not engineered into any specific layer or component, but emerge from complex interactions across the entire model. This introduces another layer of difficulty: static explanation tools that decompose models into local parts are fundamentally ill-equipped to characterize global, synergistic phenomena. When behavior is not traceable to any single input feature or neuron, how can methods based on reductionist decomposition hope to explain it?

Taken together, the intertwined challenges of sensitivity, non-linearity, and emergence point to a fundamental challenge of developing approaches to explain DFMs. They suggest that explaining DFMs may require moving beyond single-instance, deterministic explanations toward frameworks that can account for distributed, dynamic, and path-dependent behaviors—properties that lie at the heart of these complex systems.

\section*{The Barrier of Unidentifiable Data Influence}

Finally, a significant, often underestimated, barrier to explanation arises from the complex and often untraceable influence of the \emph{training data} used to create DFMs. These models learn from datasets of unprecedented scale, datasets which fundamentally shape their knowledge, capabilities, and biases. Yet, the sheer volume and heterogeneity of this data make tracing specific model behaviors back to individual influential data points exceptionally challenging. Furthermore, this tracing is profoundly complicated because data points rarely exert influence in isolation; rather, their impact is shaped through complex interactions interacting with countless other examples during the high-dimensional, non-convex optimization process. How can we ground our understanding of model behavior without a clear grasp of both the specific ``experiences'' and their intricate interplay?

This challenge of unidentifiable and interactive data influence prohibits various explainability efforts. Formal \emph{data attribution} techniques, aiming to quantify the influence of training samples on predictions, face hurdles beyond just immense computational cost and the complexity of data interactions~\cite{singh2024rethinking}. Furthermore, considering the aforementioned non-linearity and sensitivity, the complex dynamics of the training process on those data, including factors like data ordering, learning rate schedules, optimizer choices, and catastrophic forgetting, create intricate path dependencies. The influence of any single data point is mediated through these dynamics, making static attribution nearly impossible. For example, consider the fine-tuning process: early examples might steer the model significantly, but their precise influence can be overwritten or masked by later examples or by catastrophic forgetting of pre-training knowledge. Performing a proper calculation of the final contribution of any specific data point, considering the entire sequence of updates and interactions, probably represents a computationally insurmountable task, bordering on the theoretical impossibility for large-scale DFM training runs. 

Beyond specific attribution methods, this fundamental difficulty in mapping behavior back to data influence complicates the validation of \textit{any} explanation derived through other means. How can we confidently ascertain whether an explanation generated by an XAI tool reflects a genuine internal mechanism, rather than an artifact learned from obscure correlations, biases, or the untraceable residue of complex data interactions and training dynamics? Without better methods for understanding data provenance and its tangled, dynamic influence, explanations risk being contextually unmoored or potentially misleading. The difficulty in tracing interactive and path-dependent data influence thus forms a foundational barrier, challenging our ability to verify and contextualize our understanding of DFM behavior.

\section*{Implications of the Explanation Barriers}

The exploration of the intrinsic barriers confronting the explainability of DFMs – scale, non-linearity, and opaque data attribution – paints a challenging picture. If, as argued, these are not merely temporary limitations of technique but fundamental characteristics of the systems themselves, what are the necessary consequences for how we develop, deploy, and govern these powerful technologies? Recognizing these unmounted obstacles compels a reassessment of our strategies across several key domains.

Given these barriers, how then should we approach \emph{trust and reliability}? If full mechanistic understanding is fundamentally constrained by scale and emergence, relying solely on explanation methods to build trust seems insufficient. Does this necessitate a decisive shift towards validation grounded in rigorous, extensive empirical evidence? Perhaps trust must be predicated less on ``understanding the mechanism'' and more on ``demonstrating reliable behavior'' through comprehensive testing, adversarial robustness evaluations, meticulous uncertainty quantification, and transparent reporting of known limitations and failure modes.

What are the ramifications for \emph{safety, debugging, and alignment}? The non-linearity and sensitivity inherent in these models, coupled with the difficulty in tracing emergent behaviors, complicate traditional debugging. If pinpointing root causes at the parameter level is often intractable, must safety assurance and error correction pivot towards behavioral analysis? This might involve focusing on identifying failure modes through systematic testing, implementing robust input validation and output monitoring, and developing targeted interventions without necessarily comprehending the deepest underlying causes. Similarly, achieving reliable alignment with human values becomes arguably harder; if we cannot fully explain ``why'' a model behaves consistently with certain principles, reliance may increase on empirical alignment techniques like RLHF, demanding concurrent research into their own robustness and potential unintended consequences.

How does the explanation barrier impact the role of DFMs in \emph{scientific discovery}? These models excel at identifying complex patterns in data, potentially accelerating hypothesis generation. But if the process by which a DFM arrives at a novel scientific insight remains largely opaque due to its intrinsic complexity, how fully can that insight be integrated into established scientific knowledge, which typically values causal, mechanistic understanding? This necessitates developing new epistemological frameworks for evaluating and incorporating knowledge generated by inscrutable intelligence.

Finally, the intrinsic barriers pose significant questions for \emph{regulation and accountability}. How can effective governance be established for systems whose internal logic defies complete explanation? If demanding full algorithmic transparency is impractical due to inherent model properties, regulatory strategies might need to concentrate on verifiable external characteristics: setting stringent standards for performance, safety, and fairness; mandating rigorous testing and auditing protocols; requiring transparency about training data characteristics (where possible) and known model limitations; and establishing clear frameworks for risk assessment and accountability based on outcomes and processes, rather than solely on internal mechanisms.

Confronting these implications suggests critical avenues for future work. Beyond refining XAI techniques, perhaps greater emphasis should fall on developing alternative methods for assurance and control that are less dependent on full mechanistic understanding. Does the path forward involve exploring fundamentally different, potentially more constrained, AI architectures designed with interpretability as a primary objective? Or does it require cultivating a new science of complex artificial systems, focusing on characterizing their macroscopic behavior and statistical properties, accepting the limits of microscopic reductionism? Addressing the challenges posed by intrinsic barriers demands not only technical innovation but also a potential paradigm shift in how we approach the verification, validation, and governance of advanced AI.

\section*{Conclusion}

This paper has critically examined the feasibility of achieving deep, mechanistic explanations for today's Deep Foundation Models. Our exploration strongly suggests that the innate difficulties are not merely transient shortcomings of current techniques, but stem from intrinsic barriers to the very nature of these complex systems – their scale, dynamics, emergent properties, and relationship with their training data. The inherent characteristics analyzed throughout this work collectively challenge the core assumptions and capabilities of our existing explainability toolkit against DFMs.

If, as we argue, the path to full transparency is fundamentally obstructed by these intrinsic barriers, then the implications for how we engage with DFMs are significant. It compels a shift in perspective: away from potentially chasing unattainable levels of mechanistic understanding, and towards developing robust alternative frameworks for assurance. This necessitates prioritizing rigorous empirical validation, fostering new approaches to safety and alignment that can function effectively amidst opacity, and adapting our regulatory and scientific expectations. Confronting the limits of explainability calls for intellectual humility and pragmatic innovation, focusing our efforts on building trust and ensuring responsible deployment through demonstrable performance and carefully characterized limitations, rather than solely through the lens of complete explanations.

% \begin{acks}
% \end{acks}

\bibliographystyle{ACM-Reference-Format}
\bibliography{sample-base}


\begin{thebibliography}{10}

%%% the Web address.

\ifx \showCODEN    \undefined \def \showCODEN     #1{\unskip}     \fi
\ifx \showISBNx    \undefined \def \showISBNx     #1{\unskip}     \fi
\ifx \showISBNxiii \undefined \def \showISBNxiii  #1{\unskip}     \fi
\ifx \showISSN     \undefined \def \showISSN      #1{\unskip}     \fi
\ifx \showLCCN     \undefined \def \showLCCN      #1{\unskip}     \fi
\ifx \shownote     \undefined \def \shownote      #1{#1}          \fi
\ifx \showarticletitle \undefined \def \showarticletitle #1{#1}   \fi
\ifx \showURL      \undefined \def \showURL       {\relax}        \fi
% invisible to TeX
\providecommand\bibfield[2]{#2}
\providecommand\bibinfo[2]{#2}
\providecommand\natexlab[1]{#1}
\providecommand\showeprint[2][]{arXiv:#2}

\bibitem[Berglund et~al\mbox{.}(2024)]%
        {berglund2024the}
\bibfield{author}{\bibinfo{person}{Lukas Berglund}, \bibinfo{person}{Meg Tong}, \bibinfo{person}{Maximilian Kaufmann}, \bibinfo{person}{Mikita Balesni}, \bibinfo{person}{Asa~Cooper Stickland}, \bibinfo{person}{Tomasz Korbak}, {and} \bibinfo{person}{Owain Evans}.} \bibinfo{year}{2024}\natexlab{}.
\newblock \showarticletitle{The Reversal Curse: {LLM}s trained on {\textquotedblleft}A is B{\textquotedblright} fail to learn {\textquotedblleft}B is A{\textquotedblright}}. In \bibinfo{booktitle}{\emph{The Twelfth International Conference on Learning Representations}}.
\newblock
\urldef\tempurl%
\url{https://openreview.net/forum?id=GPKTIktA0k}
\showURL{%
\tempurl}


\bibitem[Casper et~al\mbox{.}(2023)]%
        {Casper2023ExplainingNeurons}
\bibfield{author}{\bibinfo{person}{Steve Casper}, \bibinfo{person}{Carson Denison}, \bibinfo{person}{Jesse Alam}, \bibinfo{person}{Nelson Elhage}, \bibinfo{person}{Nicklas Nolte}, \bibinfo{person}{Chris Olah}, \bibinfo{person}{Catherine Olsson}, \bibinfo{person}{Andy Jones}, \bibinfo{person}{Anson Ho}, \bibinfo{person}{Sam Bowman}, \bibinfo{person}{Jacob Steinhardt}, \bibinfo{person}{Dylan Hadfield-Menell}, \bibinfo{person}{Alan Coletta}, {and} \bibinfo{person}{Tomer Kaftan}.} \bibinfo{year}{2023}\natexlab{}.
\newblock \bibinfo{booktitle}{\emph{{Explaining neurons in language models}}}.
\newblock \bibinfo{type}{{T}echnical {R}eport}. \bibinfo{institution}{OpenAI}.
\newblock
\urldef\tempurl%
\url{https://openaipublic.blob.core.windows.net/neuron-explainer/paper/index.html}
\showURL{%
\tempurl}
\newblock
\shownote{Accessed: 2025-04-16}.


\bibitem[Jordan(2024)]%
        {jordan2024on}
\bibfield{author}{\bibinfo{person}{Keller Jordan}.} \bibinfo{year}{2024}\natexlab{}.
\newblock \showarticletitle{On the Variance of Neural Network Training with respect to Test Sets and Distributions}. In \bibinfo{booktitle}{\emph{The Twelfth International Conference on Learning Representations}}.
\newblock
\urldef\tempurl%
\url{https://openreview.net/forum?id=pEGSdJu52I}
\showURL{%
\tempurl}


\bibitem[Lundberg and Lee(2017)]%
        {lundberg2017unified}
\bibfield{author}{\bibinfo{person}{Scott~M Lundberg} {and} \bibinfo{person}{Su-In Lee}.} \bibinfo{year}{2017}\natexlab{}.
\newblock \showarticletitle{A unified approach to interpreting model predictions}.
\newblock \bibinfo{journal}{\emph{Advances in neural information processing systems}}  \bibinfo{volume}{30} (\bibinfo{year}{2017}).
\newblock


\bibitem[Ribeiro et~al\mbox{.}(2016)]%
        {ribeiro2016should}
\bibfield{author}{\bibinfo{person}{Marco~Tulio Ribeiro}, \bibinfo{person}{Sameer Singh}, {and} \bibinfo{person}{Carlos Guestrin}.} \bibinfo{year}{2016}\natexlab{}.
\newblock \showarticletitle{" Why should i trust you?" Explaining the predictions of any classifier}. In \bibinfo{booktitle}{\emph{Proceedings of the 22nd ACM SIGKDD international conference on knowledge discovery and data mining}}. \bibinfo{pages}{1135--1144}.
\newblock


\bibitem[Sigma(2019)]%
        {sigma2019interpretability}
\bibfield{author}{\bibinfo{person}{Two Sigma}.} \bibinfo{year}{2019}\natexlab{}.
\newblock \showarticletitle{Interpretability methods in machine learning: A brief survey}.
\newblock \bibinfo{journal}{\emph{Retrieved September}}  \bibinfo{volume}{9} (\bibinfo{year}{2019}), \bibinfo{pages}{2022}.
\newblock


\bibitem[Singh et~al\mbox{.}(2024)]%
        {singh2024rethinking}
\bibfield{author}{\bibinfo{person}{Chandan Singh}, \bibinfo{person}{Jeevana~Priya Inala}, \bibinfo{person}{Michel Galley}, \bibinfo{person}{Rich Caruana}, {and} \bibinfo{person}{Jianfeng Gao}.} \bibinfo{year}{2024}\natexlab{}.
\newblock \showarticletitle{Rethinking interpretability in the era of large language models}.
\newblock \bibinfo{journal}{\emph{arXiv preprint arXiv:2402.01761}} (\bibinfo{year}{2024}).
\newblock


\bibitem[Tan et~al\mbox{.}(2024)]%
        {tan2024interpreting}
\bibfield{author}{\bibinfo{person}{Zhen Tan}, \bibinfo{person}{Lu Cheng}, \bibinfo{person}{Song Wang}, \bibinfo{person}{Bo Yuan}, \bibinfo{person}{Jundong Li}, {and} \bibinfo{person}{Huan Liu}.} \bibinfo{year}{2024}\natexlab{}.
\newblock \showarticletitle{Interpreting pretrained language models via concept bottlenecks}. In \bibinfo{booktitle}{\emph{Pacific-Asia Conference on Knowledge Discovery and Data Mining}}. Springer, \bibinfo{pages}{56--74}.
\newblock


\bibitem[Wei et~al\mbox{.}(2022)]%
        {wei2022emergent}
\bibfield{author}{\bibinfo{person}{Jason Wei}, \bibinfo{person}{Yi Tay}, \bibinfo{person}{Rishi Bommasani}, \bibinfo{person}{Colin Raffel}, \bibinfo{person}{Barret Zoph}, \bibinfo{person}{Sebastian Borgeaud}, \bibinfo{person}{Dani Yogatama}, \bibinfo{person}{Maarten Bosma}, \bibinfo{person}{Denny Zhou}, \bibinfo{person}{Donald Metzler}, \bibinfo{person}{Ed~H. Chi}, \bibinfo{person}{Tatsunori Hashimoto}, \bibinfo{person}{Oriol Vinyals}, \bibinfo{person}{Percy Liang}, \bibinfo{person}{Jeff Dean}, {and} \bibinfo{person}{William Fedus}.} \bibinfo{year}{2022}\natexlab{}.
\newblock \showarticletitle{Emergent Abilities of Large Language Models}.
\newblock \bibinfo{journal}{\emph{Transactions on Machine Learning Research}} (\bibinfo{year}{2022}).
\newblock
\showISSN{2835-8856}
\urldef\tempurl%
\url{https://openreview.net/forum?id=yzkSU5zdwD}
\showURL{%
\tempurl}
\newblock
\shownote{Survey Certification}.


\bibitem[Zhou et~al\mbox{.}(2024)]%
        {zhou2024comprehensive}
\bibfield{author}{\bibinfo{person}{Ce Zhou}, \bibinfo{person}{Qian Li}, \bibinfo{person}{Chen Li}, \bibinfo{person}{Jun Yu}, \bibinfo{person}{Yixin Liu}, \bibinfo{person}{Guangjing Wang}, \bibinfo{person}{Kai Zhang}, \bibinfo{person}{Cheng Ji}, \bibinfo{person}{Qiben Yan}, \bibinfo{person}{Lifang He}, {et~al\mbox{.}}} \bibinfo{year}{2024}\natexlab{}.
\newblock \showarticletitle{A comprehensive survey on pretrained foundation models: A history from bert to chatgpt}.
\newblock \bibinfo{journal}{\emph{International Journal of Machine Learning and Cybernetics}} (\bibinfo{year}{2024}), \bibinfo{pages}{1--65}.
\newblock


\end{thebibliography}

% \appendix

\end{document}